\documentclass[sigconf]{acmart}

\setcopyright{none}
\renewcommand\footnotetextcopyrightpermission[1]{}
\usepackage{booktabs}
\usepackage{subcaption}
\usepackage{placeins} 
\usepackage{enumitem} 

\title[Do Deployment Constraints Make LLMs Hallucinate Citations?]{Do Deployment Constraints Make LLMs Hallucinate Citations? An Empirical Study across Four Models and Five Prompting Regimes}

\author{Chen Zhao}
\email{cz1296@nyu.edu}
\affiliation{%
  \institution{New York University}
  \country{United States}
}

\author{Yuan Tang}
\email{yuantang@alumni.cmu.edu}
\affiliation{%
  \institution{Carnegie Mellon University}
  \country{United States}
}

\author{Yitian Qian}
\email{qyt024@bu.edu}
\affiliation{%
  \institution{Boston University}
  \country{United States}
}

\begin{document}

\begin{abstract}
LLMs are increasingly used to draft academic text and to support software engineering (SE) evidence synthesis, but they often hallucinate bibliographic references that look legitimate. We study how deployment-motivated prompting constraints affect citation verifiability in a closed-book setting. Using 144 claims (24 in SE~\&~CS) and a deterministic verification pipeline (Crossref + Semantic Scholar), we evaluate two proprietary models (Claude Sonnet, GPT-4o) and two open-weight models (LLaMA~3.1--8B, Qwen~2.5--14B) across five regimes: Baseline, Temporal (publication-year window), Survey-style breadth, Non-Disclosure policy, and their combination. Across 17{,}443 generated citations, no model exceeds a \emph{citation-level} existence rate of 0.475; Temporal and Combo conditions produce the steepest drops while outputs remain format-compliant (well-formed bibliographic fields). Unresolved outcomes dominate (36--61\%); a 100-citation audit indicates that a substantial fraction of Unresolved cases are fabricated. Results motivate post-hoc citation verification before LLM outputs enter SE literature reviews or tooling pipelines.
\end{abstract}



\maketitle

\section{Introduction}
Large language models (LLMs) have demonstrated strong capabilities in generating fluent academic prose, leading to their increasing use in scholarly writing and evidence synthesis. Yet they continue to exhibit \emph{citation hallucination}~\cite{walters2023fabrication,agrawal2024know}: the generation of bibliographic references that look complete (authors, venues, DOIs) but do not correspond to verifiable works. These errors are often attributed to next-token prediction objectives and distributional shifts~\cite{mckenna2023sources}. However, how citation reliability behaves under realistic deployment constraints---temporal restrictions, survey-style breadth pressure, and \emph{non-disclosure} policies that forbid claims of training-data access---remains poorly understood.

This problem is especially relevant to software engineering (SE) research, where systematic literature reviews (SLRs), mapping studies, and related-work sections are core methodological activities~\cite{kitchenham2007slr}. As LLMs are increasingly integrated into SE workflows---from drafting technical reports and architecture decision records to assisting with literature surveys~\cite{fan2023llmse,hou2024llmse}---unreliable citations can propagate through evidence synthesis pipelines that the SE community relies on. Understanding how deployment constraints affect citation quality is therefore a prerequisite for responsibly adopting LLM-assisted writing tools in SE practice.

We address this gap with an automated verification framework that parses each citation, queries Crossref and Semantic Scholar, and assigns one of three labels: \emph{Existing}, \emph{Unresolved}, or \emph{Fabricated}. We evaluate four LLMs under five prompting conditions with deterministic decoding and find that temporal constraints are associated with the steepest drop in verifiability even as models maintain format compliance; survey-style prompting widens the proprietary--open-weight gap in verifiability; and unresolved outcomes constitute 36--61\% of citations, forming the largest category in the majority of model--condition cells.

We structure our investigation around three research questions:
\begin{description}[nosep,leftmargin=1em,labelindent=0em]
\item[RQ1] How do deployment constraints (temporal, survey-style, non-disclosure) affect citation verifiability?
\item[RQ2] How do these effects differ between proprietary and open-weight models?
\item[RQ3] How does combining all three constraints compare to applying each individually?
\end{description}

Our contributions are (1) a curated dataset of 144 claims spanning six academic domains (including 24 in SE~\&~CS), (2) an automated verification pipeline with an audited three-way label taxonomy (Cohen's $\kappa{=}0.63$ vs.\ human labels), and (3) an empirical analysis of how deployment constraints and the proprietary--open-weight distinction affect citation verifiability.

\section{Related Work}

\paragraph{Hallucination in language models.}
LLM hallucination has been widely studied across summarization and knowledge-intensive generation~\cite{maynez2020faithfulness,huang2023survey,tonmoy2024survey,alansari2025hallucination_survey}.
\citet{mckenna2023sources} trace many failures to distributional gaps between training and deployment, while \citet{kalai2025why} argue that next-token prediction can incentivize plausible completion over calibrated uncertainty.
We study an audit-friendly manifestation: hallucinated bibliographic citations, where outputs can be checked against external scholarly indexes.

\paragraph{Citation hallucination and verifiable generation.}
Prior work audits LLM-generated references but often adopts a binary real-or-fabricated label under a single prompting setup.
\citet{walters2023fabrication} reported high rates of non-existent citations from ChatGPT, and \citet{agrawal2024know} showed that hallucinated references often contain internally inconsistent metadata.
Complementary work evaluates cited generation more broadly: ALCE benchmarks end-to-end generation with citations~\cite{gao2023enabling}, and \citet{liu2023evaluating} audit citation precision and recall in generative search engines.
Our work systematically varies \emph{deployment constraints} (Temporal, Survey-style breadth, Non-Disclosure, and Combo) in a closed-book reference-list setting and uses a three-way taxonomy (\emph{Existing}, \emph{Unresolved}, \emph{Fabricated}) that distinguishes clearly fabricated references from cases that cannot be reliably verified.

\paragraph{LLMs for scientific writing.}
Retrieval-augmented pipelines with structured planning can reduce hallucination in literature review generation~\cite{agarwal2024litllm}, yet hallucinated references remain common even with explicit grounding instructions~\cite{cai2025litreview}.
These issues are particularly salient for evidence-synthesis workflows (e.g., SLRs) that are common in SE research~\cite{kitchenham2007slr} and for emerging LLM-assisted SE tooling surveyed in recent work~\cite{fan2023llmse,hou2024llmse}.
Our work is complementary: rather than proposing a new generation pipeline, we present a deterministic verification framework and use it to quantify how citation reliability changes under deployment-motivated prompting constraints in a closed-book setting.

\section{Experimental Design}
\label{sec:setup}

\paragraph{Task formulation.}
We study citation reliability in LLM-generated academic-style writing. Each model receives a question-style prompt and generates a concise academic paragraph followed by a structured reference list with fixed bibliographic fields (title, authors, venue, year, DOI, and URL when available), enabling deterministic parsing and automated verification. Prompts request exactly $k$ citations (5 for Baseline/Temporal/Non-Disclosure; 8 for Survey/Combo), but we do not enforce count compliance: all references in the output are verified and the realized count is recorded. Verification uses a weighted metadata similarity score (Eq.~\ref{eq:score}); details are in Section~\ref{sec:pipeline}.

\paragraph{Claim dataset.}
Our dataset contains 144 question-style prompts (e.g., ``What evidence supports\ldots'') spanning six domain groups: SE~\&~CS (24 claims covering software engineering, AI/ML, data/HCI, security, and systems), Natural Sciences (24), Medicine~\&~Health (24), Social Sciences (24), Humanities (24), and Interdisciplinary (24). Each record includes a domain label, an optional temporal window, and optional seed anchors (keyword-only topic hints that orient the model without naming specific papers). When present, the same anchors are provided across all conditions for a given claim. The 144 prompts were randomly sampled (fixed seed) from a candidate pool of 240 items sourced from publicly accessible academic materials, pre-screened for suitability, and stratified by domain.

\paragraph{Models.}
We evaluate two proprietary models---Claude~Sonnet (Anthropic) and GPT-4o (OpenAI)---and two open-weight models---Qwen~2.5--14B (Alibaba) and LLaMA~3.1--8B (Meta)~\cite{anthropic2024claude_docs,openai2024gpt4o_systemcard,yang2024qwen25,meta2024llama31}.\footnote{Exact model identifiers are listed in the replication package.}
Because the proprietary models are presumably larger, our design cannot fully disentangle model scale from the proprietary--open-weight distinction (Section~\ref{sec:discussion}).

\paragraph{Prompting conditions.}
Each claim is evaluated under five conditions via fixed prompt templates. \emph{Baseline} requests one academic paragraph with exactly 5 citations. \emph{Temporal} adds a strict publication-year window (5 citations); all 144 windows end at 2025 with a median span of five years, targeting recent literature where parametric knowledge is weakest. \emph{Survey} asks for a related-work synthesis organized into 3--4 approach categories with 8 citations, reflecting breadth pressure in evidence-synthesis writing. Survey/Combo request 8 citations (vs.\ 5 for the other conditions) because a 5-citation budget yields fewer than two references per category, which is atypical for survey-style writing; since our primary outcomes are citation-level proportions, the differing budgets change observation counts rather than outcome definitions, and we additionally report per-claim verification fractions (Figure~\ref{fig:frac_boxplot}) for a claim-equal view. \emph{Non-Disclosure} instructs the model not to claim access to memorized training documents (5 citations), simulating deployment policies in commercial writing assistants. \emph{Combo} combines all three constraints (8 citations). All conditions use deterministic decoding (temperature~$0$) with no retrieval augmentation. The full design yields $144 \times 5 \times 4 = 2{,}880$ runs producing 17{,}443 individual citations.

\section{Verification Pipeline}\label{sec:pipeline}

Our pipeline checks every generated citation against two scholarly databases---Crossref and Semantic Scholar---which together cover most of the indexed literature. The same pipeline runs identically on every model and condition. The code is publicly available.\footnote{See the replication package at \url{https://github.com/Zerichen/Citation-Hallucination}.}

\paragraph{Parsing.}
The pipeline extracts the citation list from each model output and breaks every reference into structured fields: title, authors, venue, year, and DOI. As in the task formulation, all references present in the output are extracted (no truncation, padding, or retries). We use rule-based heuristics and regular expressions; references that cannot be parsed are carried forward as unresolved.

\paragraph{Candidate retrieval.}
For each parsed citation, we run up to three lookups: (1) if the model supplied a DOI, we check it directly via Crossref; (2) we search Semantic Scholar by title (returning up to $k{=}5$ candidates); (3) we search Crossref by title (again up to $k{=}5$). All retrieved records are normalized into a common format.

\paragraph{Scoring.}
Each candidate is scored against the parsed citation using a weighted combination:
\begin{equation}\label{eq:score}
s = 0.60 \cdot t + 0.20 \cdot a + 0.15 \cdot y + 0.05 \cdot v
\end{equation}
where $t$ is fuzzy title similarity (token-set ratio), $a$ is author last-name overlap, $y$ is year agreement (1.0 if exact, 0.5 if off by one, 0 otherwise), and $v$ is venue similarity (partial ratio). Title receives the heaviest weight because it is the most discriminative field in practice.

\paragraph{Labeling.}
Each citation receives one of three labels: \emph{Existing} (score $\geq 0.85$), \emph{Unresolved} ($0.60 \leq$ score $< 0.85$), or \emph{Fabricated} (score $< 0.60$ or no candidate found). For temporal conditions, citations outside the stated publication-year window are additionally flagged. We set $0.85$ as the match threshold: above this score, candidates consistently correspond to the correct paper despite minor metadata drift (e.g., abbreviated venues, missing middle initials). The $0.60$ boundary separates weak, likely spurious matches from partially plausible ones; below $0.60$, candidates typically share only generic tokens. \emph{Unresolved} is intentionally a triage bucket: it includes (i) real papers the pipeline cannot fully confirm due to incomplete or discordant metadata, and (ii) fabricated citations that partially overlap with unrelated records. We therefore report it separately rather than force a binary decision.

\paragraph{Aggregation and metrics.}
Rates of \emph{Existing}, \emph{Fabricated}, and \emph{Unresolved} (summing to 1) are averaged over all citations in a model--condition cell, with 95\% cluster-bootstrap CIs (1{,}000 resamples over 144 claims) to account for within-claim correlation. For key pairwise comparisons, we report the bootstrap CI of the \emph{difference} ($\Delta$) in existence rate~\cite{schenker2001overlap}; a difference is statistically meaningful when its CI excludes zero. We complement citation-weighted rates with the \emph{per-claim verification fraction} (Figure~\ref{fig:frac_boxplot}), $f_i=\#\emph{Existing}_i / \#\emph{TotalParsed}_i$, which weights each claim equally regardless of citation count.

\paragraph{Pipeline validation.}
We manually verified a stratified sample of 100 citations against Google Scholar and DBLP (Table~\ref{tab:confusion}). Overall agreement was 75\% and Cohen's $\kappa$ (pipeline vs.\ human) was 0.63. Precision was 0.97 for \emph{Existing} and 0.88 for \emph{Fabricated}, but only 0.43 for \emph{Unresolved}. The dominant error was \emph{Unresolved}~$\to$ \emph{Fabricated} (16 cases), so our reported fabrication rates are likely conservative; we therefore treat \emph{Unresolved} as high-risk and report a reclassification sensitivity analysis in Section~\ref{sec:results}.

\begin{table}[t!]
\small
\caption{Pipeline vs.\ human labels on a stratified sample of 100 citations. Overall agreement: 75\%; Cohen's $\kappa{=}0.63$.}
\label{tab:confusion}
\begin{tabular}{@{}lccc@{}}
\toprule
 & \multicolumn{3}{c}{\textbf{Human label}} \\
\cmidrule(l){2-4}
\textbf{Pipeline label} & Exist. & Unres. & Fabric. \\
\midrule
Existing    & 31 & 1  &  0 \\
Unresolved  &  4 & 15 & 16 \\
Fabricated  &  2 &  2 & 29 \\
\bottomrule
\end{tabular}
\end{table}

\begin{table*}[t]
\caption{Citation-level verification metrics with 95\% bootstrap CIs ($N{=}144$ claims per cell). Conditions: Base = Baseline, Temp = Temporal, Surv = Survey, Non-Disc.\ = Non-Disclosure. T.~Viol.\ = temporal violation rate.}
\label{tab:citation_metrics}
\footnotesize
\setlength{\tabcolsep}{3.5pt}
\renewcommand{\arraystretch}{0.92}
\begin{tabular}{ll r ccc rr}
\toprule
\textbf{Model} & \textbf{Cond.} & \textbf{N} & \textbf{Existing} $\uparrow$ & \textbf{Fabricated} $\downarrow$ & \textbf{Unresolved} & \textbf{T.~Viol.} & \textbf{Avg.~\#Cit.} \\
\midrule
Claude Sonnet  & Base  & 144 & .381\,{\scriptsize[.335,.426]} & .157\,{\scriptsize[.126,.190]} & .462\,{\scriptsize[.382,.549]} & .000 & 5.00 \\
GPT-4o         & Base  & 144 & .235\,{\scriptsize[.189,.281]} & .281\,{\scriptsize[.239,.327]} & .484\,{\scriptsize[.403,.569]} & .000 & 4.97 \\
LLaMA~3.1--8B  & Base  & 144 & .068\,{\scriptsize[.049,.090]} & .369\,{\scriptsize[.327,.411]} & .563\,{\scriptsize[.479,.646]} & .000 & 4.98 \\
Qwen~2.5--14B  & Base  & 144 & .090\,{\scriptsize[.065,.118]} & .442\,{\scriptsize[.398,.486]} & .468\,{\scriptsize[.382,.549]} & .000 & 4.98 \\
\midrule
Claude Sonnet  & Temp  & 144 & .119\,{\scriptsize[.089,.151]} & .347\,{\scriptsize[.306,.392]} & .533\,{\scriptsize[.451,.618]} & .015 & 5.00 \\
GPT-4o         & Temp  & 144 & .019\,{\scriptsize[.008,.033]} & .451\,{\scriptsize[.412,.493]} & .529\,{\scriptsize[.444,.611]} & .001 & 5.00 \\
LLaMA~3.1--8B  & Temp  & 144 & .011\,{\scriptsize[.004,.019]} & .394\,{\scriptsize[.351,.440]} & .595\,{\scriptsize[.514,.674]} & .026 & 5.01 \\
Qwen~2.5--14B  & Temp  & 144 & .014\,{\scriptsize[.004,.025]} & .455\,{\scriptsize[.408,.499]} & .531\,{\scriptsize[.444,.611]} & .015 & 4.99 \\
\midrule
Claude Sonnet  & Surv  & 144 & .475\,{\scriptsize[.425,.523]} & .161\,{\scriptsize[.133,.189]} & .364\,{\scriptsize[.285,.444]} & .000 & 8.00 \\
GPT-4o         & Surv  & 144 & .203\,{\scriptsize[.165,.246]} & .302\,{\scriptsize[.262,.345]} & .495\,{\scriptsize[.410,.576]} & .000 & 7.59 \\
LLaMA~3.1--8B  & Surv  & 144 & .038\,{\scriptsize[.025,.053]} & .436\,{\scriptsize[.392,.475]} & .526\,{\scriptsize[.444,.611]} & .000 & 7.95 \\
Qwen~2.5--14B  & Surv  & 144 & .020\,{\scriptsize[.009,.032]} & .547\,{\scriptsize[.509,.584]} & .433\,{\scriptsize[.347,.514]} & .000 & 7.28 \\
\midrule
Claude Sonnet  & Non-Disc.\ & 144 & .349\,{\scriptsize[.302,.397]} & .165\,{\scriptsize[.129,.201]} & .487\,{\scriptsize[.403,.569]} & .000 & 4.90 \\
GPT-4o         & Non-Disc.\ & 144 & .175\,{\scriptsize[.142,.210]} & .317\,{\scriptsize[.271,.360]} & .508\,{\scriptsize[.424,.590]} & .000 & 5.00 \\
LLaMA~3.1--8B  & Non-Disc.\ & 144 & .045\,{\scriptsize[.029,.066]} & .398\,{\scriptsize[.357,.435]} & .557\,{\scriptsize[.472,.639]} & .000 & 4.97 \\
Qwen~2.5--14B  & Non-Disc.\ & 144 & .078\,{\scriptsize[.054,.104]} & .410\,{\scriptsize[.368,.453]} & .512\,{\scriptsize[.431,.597]} & .000 & 4.99 \\
\midrule
Claude Sonnet  & Combo & 144 & .106\,{\scriptsize[.078,.142]} & .359\,{\scriptsize[.318,.399]} & .536\,{\scriptsize[.451,.618]} & .027 & 7.61 \\
GPT-4o         & Combo & 144 & .005\,{\scriptsize[.000,.012]} & .452\,{\scriptsize[.418,.489]} & .543\,{\scriptsize[.465,.625]} & .000 & 7.38 \\
LLaMA~3.1--8B  & Combo & 144 & .008\,{\scriptsize[.003,.014]} & .386\,{\scriptsize[.343,.423]} & .606\,{\scriptsize[.521,.681]} & .000 & 7.99 \\
Qwen~2.5--14B  & Combo & 144 & .001\,{\scriptsize[.000,.003]} & .507\,{\scriptsize[.468,.543]} & .492\,{\scriptsize[.410,.576]} & .000 & 7.56 \\
\bottomrule
\end{tabular}
\end{table*}

\section{Results}
\label{sec:results}

\begin{figure*}[t!]
    \centering
    \includegraphics[width=0.88\textwidth]{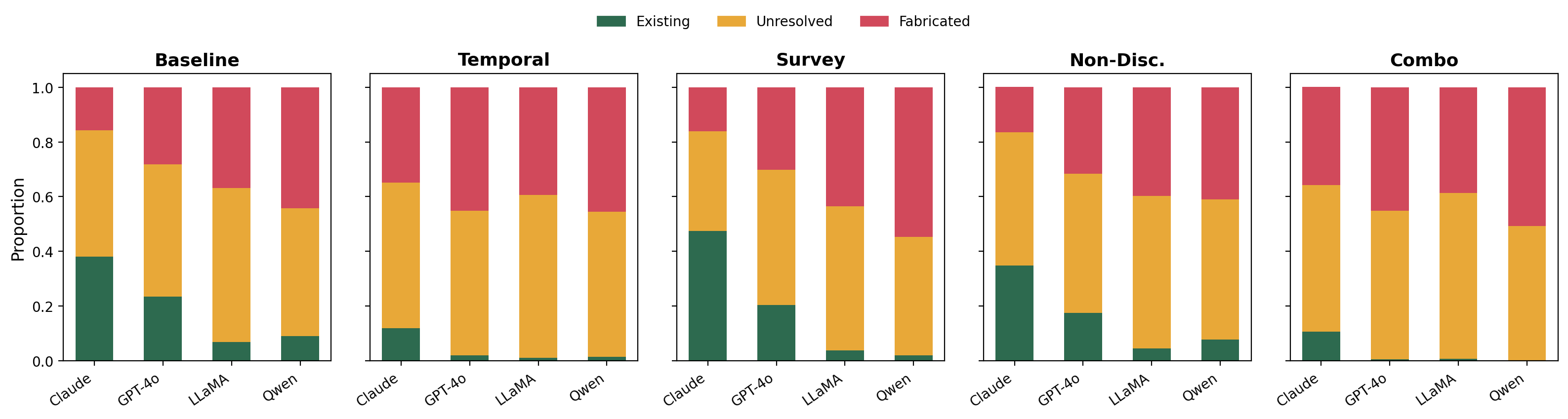}
    \vspace{-0.3em}
    \caption{Citation-level outcome distribution (Existing, Unresolved, Fabricated) for each model under all five conditions, shown as stacked proportions summing to one. ``Non-Disc.''\ denotes the non-disclosure condition.}
    \Description{Five side-by-side panels (Baseline, Temporal, Survey, Non-Disclosure, Combo). Each panel contains four vertical stacked bars (Claude, GPT-4o, LLaMA~3.1--8B, Qwen~2.5--14B) showing the proportion of citations labeled Existing, Unresolved, or Fabricated, summing to 1. Existing is highest for Claude (especially in Survey) and drops to near zero for GPT-4o and both open-weight models under Temporal and Combo, where most mass shifts to Unresolved and Fabricated.}
    \label{fig:stacked}
\end{figure*}

\begin{figure*}[t!]
    \centering
    \includegraphics[width=0.92\textwidth]{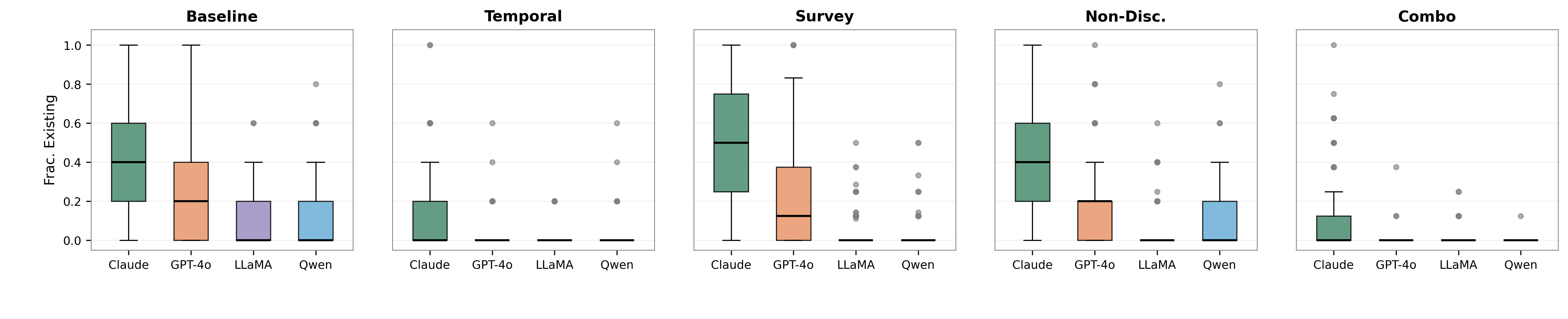}
    \vspace{-1em}
    \caption{Per-claim verification fraction by model and condition (boxes show IQR with median; whiskers follow the 1.5\,$\times$\,IQR convention). ``Non-Disc.''\ denotes the non-disclosure condition.}
    \Description{Five panels of box-and-whisker plots, one per condition (Baseline, Temporal, Survey, Non-Disc., Combo). In each panel, the x-axis lists four models (Claude, GPT-4o, LLaMA 3.1--8B, Qwen 2.5--14B) and the y-axis shows the per-claim fraction of citations verified as Existing (0 to 1). Boxes indicate the interquartile range with the median; whiskers follow the 1.5×IQR convention and outliers are shown as points. Medians are highest for Claude in Baseline and Survey and are near zero for the open-weight models in most stressed conditions, especially Temporal and Combo.}
    \label{fig:frac_boxplot}
\end{figure*}

\begin{figure}[t!]
    \centering
    \includegraphics[width=\columnwidth]{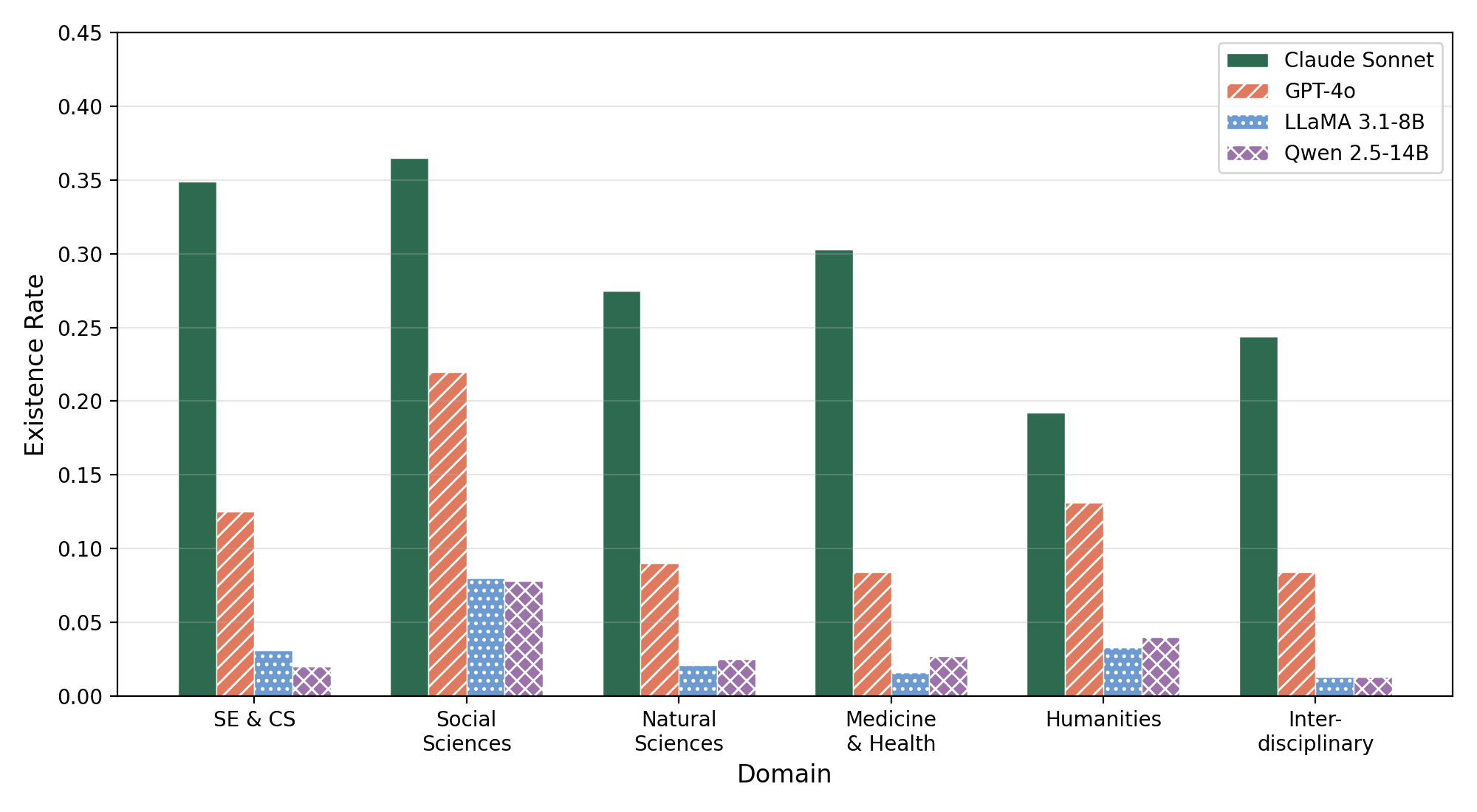}
    \vspace{-1em}
    \caption{Existence rate by domain (24~claims per group, aggregated across conditions).}
    \label{fig:domain}
\end{figure}

Table~\ref{tab:citation_metrics} reports citation-weighted verification metrics; Figure~\ref{fig:stacked} visualizes the outcome distribution; Figure~\ref{fig:frac_boxplot} summarizes per-claim verification fractions (weighting each claim equally regardless of citation count).
Figure~\ref{fig:domain} breaks down existence rate by domain; Table~\ref{tab:delta} provides bootstrap CIs of the difference ($\Delta$) in existence rate for key pairwise comparisons.

\paragraph{No model verifies a majority of its citations.}
No model, under any condition, achieves an existence rate above 0.50 (the peak is 0.475 for Claude Sonnet under Survey).
Under the strictest interpretation---counting a claim as verified only if \emph{every} citation passes---very few claims qualify even for the best model.
Different constraints produce different failure signatures, not just uniform degradation.

\paragraph{Per-claim distributions reveal partial verification.}
Figure~\ref{fig:frac_boxplot} shows the per-claim verification fraction. Even when aggregate rates are low, some claims receive partially verified sets: Claude Sonnet at Baseline has a median per-claim fraction of 0.40 (IQR 0.20--0.60), while for open-weight models the median is zero in most conditions. Low aggregates do not mean every claim is entirely unverified, but many claims receive no verified citations at all.

\paragraph{Proprietary models do better, but not well (RQ2).}
At baseline, Claude Sonnet and GPT-4o achieve existence rates of 0.381 and 0.235, while LLaMA~3.1--8B and Qwen~2.5--14B reach only 0.068 and 0.090.
The proprietary--open-weight gap is large and statistically clear ($\Delta = +0.229$, 95\% CI $[0.191, 0.266]$; Table~\ref{tab:delta}), not explained by citation volume (${\approx}\,5$ per claim for all models) and confirmed across the full per-claim distribution (Figure~\ref{fig:frac_boxplot}).
This difference persists and is often larger under stress, which may reflect differences in training data coverage and model scale, though we cannot isolate these factors with the current study design.

\begin{table}[t!]
\caption{Bootstrap 95\% CI of the difference ($\Delta$) in existence rate for key pairwise comparisons. Positive $\Delta$ favors the first term. CIs excluding zero are bolded; (ns) = CI overlaps zero.}
\label{tab:delta}
\footnotesize
\setlength{\tabcolsep}{3pt}
\renewcommand{\arraystretch}{0.88}
\begin{tabular}{@{}lcc@{}}
\toprule
\textbf{Comparison} & $\boldsymbol{\Delta}$ & \textbf{95\% CI} \\
\midrule
\multicolumn{3}{@{}l}{\emph{Prop.\ vs.\ open-wt.\ (exist.\ rate)}} \\
\textbf{Prop.\ vs.\ Open-wt.\ (Base)} & $\boldsymbol{+.229}$ & $[.191,\,.266]$ \\
\textbf{Prop.\ vs.\ Open-wt.\ (Temp)} & $\boldsymbol{+.057}$ & $[.037,\,.078]$ \\
\textbf{Prop.\ vs.\ Open-wt.\ (Surv)} & $\boldsymbol{+.310}$ & $[.274,\,.349]$ \\
\textbf{Prop.\ vs.\ Open-wt.\ (N-D)} & $\boldsymbol{+.200}$ & $[.164,\,.234]$ \\
\textbf{Prop.\ vs.\ Open-wt.\ (Combo)} & $\boldsymbol{+.051}$ & $[.035,\,.069]$ \\
\midrule
\multicolumn{3}{@{}l}{\emph{Constraint vs.\ Base (Claude Sonnet)}} \\
\textbf{Temp $-$ Base} & $\boldsymbol{-.261}$ & $[-.317,\,-.207]$ \\
\textbf{Surv $-$ Base} & $\boldsymbol{+.094}$ & $[.028,\,.162]$ \\
N-D $-$ Base \textit{(ns)}   & $-.032$ & $[-.095,\,.031]$ \\
\textbf{Combo $-$ Base} & $\boldsymbol{-.275}$ & $[-.329,\,-.220]$ \\
\midrule
\multicolumn{3}{@{}l}{\emph{Constraint vs.\ Base (GPT-4o)}} \\
\textbf{Temp $-$ Base} & $\boldsymbol{-.216}$ & $[-.266,\,-.168]$ \\
Surv $-$ Base \textit{(ns)}    & $-.032$ & $[-.092,\,.029]$ \\
\textbf{N-D $-$ Base} & $\boldsymbol{-.060}$ & $[-.119,\,-.001]$ \\
\textbf{Combo $-$ Base} & $\boldsymbol{-.230}$ & $[-.274,\,-.185]$ \\
\midrule
\multicolumn{3}{@{}l}{\emph{Constraint vs.\ Base (LLaMA~3.1--8B)}} \\
\textbf{Temp $-$ Base} & $\boldsymbol{-.057}$ & $[-.078,\,-.038]$ \\
\textbf{Surv $-$ Base} & $\boldsymbol{-.030}$ & $[-.048,\,-.013]$ \\
\textbf{N-D $-$ Base} & $\boldsymbol{-.023}$ & $[-.045,\,-.001]$ \\
\textbf{Combo $-$ Base} & $\boldsymbol{-.060}$ & $[-.083,\,-.040]$ \\
\midrule
\multicolumn{3}{@{}l}{\emph{Constraint vs.\ Base (Qwen~2.5--14B)}} \\
\textbf{Temp $-$ Base} & $\boldsymbol{-.076}$ & $[-.106,\,-.047]$ \\
\textbf{Surv $-$ Base} & $\boldsymbol{-.070}$ & $[-.101,\,-.041]$ \\
N-D $-$ Base \textit{(ns)}   & $-.012$ & $[-.050,\,.025]$ \\
\textbf{Combo $-$ Base} & $\boldsymbol{-.089}$ & $[-.117,\,-.063]$ \\
\bottomrule
\end{tabular}
\end{table}

\paragraph{Temporal constraints produce the steepest decline (RQ1).}
Temporal constraints reduce citation quality more than any other single condition.
Claude Sonnet falls from 0.381 to 0.119 ($\Delta = -0.261$, 95\% CI $[-0.317, -0.207]$); GPT-4o drops comparably ($\Delta = -0.216$, CI $[-0.266, -0.168]$); the open-weight models, already near zero, decline further ($\Delta = -0.076$ for Qwen and $-0.057$ for LLaMA; Table~\ref{tab:delta}).
Critically, direct temporal violations are extremely rare (0.001--0.026): the models \emph{obey} the temporal constraint but cannot produce verifiable references within it, a failure mode that format-level compliance checks would miss entirely.

\paragraph{Survey prompting is associated with a larger proprietary--open-weight gap (RQ1, RQ2).}
Under the Survey condition, the proprietary--open-weight gap increases to its largest observed value ($\Delta = +0.310$, 95\%~CI $[0.274, 0.349]$; Table~\ref{tab:delta}).
Models slightly under-produce relative to the requested 8~citations (Avg.~\#Cit.\ 7.28--8.00; Table~\ref{tab:citation_metrics}), but the pattern persists.
Claude~Sonnet \emph{improves}---existence rises from 0.381 to 0.475 ($\Delta = +0.094$, CI $[0.028, 0.162]$) with comparable fabrication (0.157$\to$0.161)---suggesting it draws on a sufficient stock of real references.
GPT-4o's decline under Survey is not statistically significant ($\Delta = -0.032$, CI $[-0.092, 0.029]$).
Open-weight models show the opposite trend: Qwen~2.5--14B drops significantly ($\Delta = -0.070$, CI $[-0.101, -0.041]$) and reaches 0.547 fabrication, the highest in the study.

\paragraph{Non-Disclosure instructions redistribute rather than eliminate errors (RQ1).}
Non-Disclosure prompting has a subtler effect: existence-rate declines are small, with marginal decreases for GPT-4o ($\Delta = -0.060$, CI $[-0.119, -0.001]$) and LLaMA ($\Delta = -0.023$, CI $[-0.045, -0.001]$), while Claude Sonnet and Qwen show no significant change (Table~\ref{tab:delta}).
What changes is the error balance: some verified citations shift into the unresolved bin (e.g., Claude Sonnet existence 0.381$\to$0.349, unresolved 0.462$\to$0.487). DOI completeness drops under Non-Disclosure (most pronounced for LLaMA, $-11.4$~percentage points\footnote{DOI completeness rates by model and condition are reported in the replication package.}), and because DOI is the strongest verification signal, its suppression pushes citations from Existing into Unresolved---shifting errors from ``obviously wrong'' into ``hard to tell.''

\paragraph{Combining constraints produces the worst outcomes (RQ3).}
{\tolerance=800 The Combo condition produces the most severe degradation (Figure~\ref{fig:stacked}).
Three of four models show existence rates near zero; only Claude~Sonnet retains a non-trivial 0.106 ($\Delta\!=\!{-}0.275$, CI~$[{-}0.329,\allowbreak {-}0.220]$).
GPT-4o drops comparably ($\Delta\!=\!{-}0.230$, CI~$[{-}0.274,\allowbreak {-}0.185]$); both open-weight models collapse (Table~\ref{tab:delta}).
For proprietary models, the combined decline exceeds the largest single-constraint drop; open-weight models exhibit a floor effect.
Critically, citation volume goes \emph{up} (7.4--8.0 per claim): models keep generating references even as verifiability erodes.\par}

\paragraph{A large fraction of citations remain automatically unresolvable.}
Unresolved outcomes account for 36--61\% of citations across nearly all cells. Manual validation shows that nearly half of sampled unresolved citations are fabricated (16 of 35; Table~\ref{tab:confusion}), so this category should not be read as ``nearly correct.'' This validates the three-way classification: collapsing into a binary scheme would hide the large pool of genuinely uncertain citations.

\paragraph{Sensitivity to reclassifying \emph{Unresolved}.}
In the audit, only 15/35 ``Unresolved'' citations were truly unresolved; 16/35 were fabricated and 4/35 were existing (Table~\ref{tab:confusion}). If we proportionally reassign each cell's Unresolved mass using these rates, fabricated rates increase to 0.33--0.75 and existence rates to 0.06--0.52, while the direction of all constraint effects and the proprietary--open-weight gap remain unchanged. This suggests our main conclusions are robust but that absolute fabrication rates are likely underestimates.

\paragraph{Domain-stratified analysis highlights SE relevance.}
Figure~\ref{fig:domain} breaks down the existence rate by domain group and model, aggregated across all five conditions; domain-level differences should be interpreted descriptively.
The SE~\&~CS group (24~claims, 2{,}926~citations) exhibits an existence rate of 0.132, comparable to the cross-domain average (0.120). Social Sciences achieves the highest rate (0.187), possibly reflecting stronger database coverage.
The proprietary--open-weight gap reproduces within every domain group: Claude Sonnet reaches 0.349 in SE~\&~CS while both open-weight models remain below 0.10.
These results suggest that the hallucination patterns documented in this study apply directly to the kind of literature SE researchers produce and consume, though subdomain comparisons should be interpreted cautiously given sample size (Section~\ref{sec:discussion}).

\section{Discussion}
\label{sec:discussion}

Citation hallucination is a systematic failure whose shape depends on the constraints placed on the model. Three themes emerge.
First, \emph{compliance without substance}: under the Temporal condition every model produces well-formed bibliographic entries that respect the requested year window, yet existence rates fall sharply---GPT-4o drops from 0.235 to 0.019, and both open-weight models reach effectively zero. Format compliance therefore masks a near-complete loss of verifiability.
Second, \emph{the unresolved-citation problem}: 36--61\% of citations across nearly all cells cannot be automatically confirmed or refuted, and our audit shows that roughly half of these are fabricated (Table~\ref{tab:confusion}). Any binary real-or-fabricated evaluation would hide this large, high-risk category.
Third, the \emph{proprietary--open-weight gap} ($\Delta$ up to $+0.310$; Table~\ref{tab:delta}) persists across all conditions, suggesting that the proprietary--open-weight distinction may be a stronger predictor of citation quality than any single prompting constraint.

\subsection{Qualitative Error Patterns}

Manual inspection of fabricated and unresolved citations reveals four recurring failure modes that format-level checks would not catch:

\begin{itemize}[nosep,leftmargin=1.2em]
\item \textbf{Venue laundering.} A plausible venue name (e.g., \emph{IEEE TSE}, \emph{ICSE}) is paired with a title that does not appear in that venue's index---or in any index.
\item \textbf{Author bricolage.} Real-sounding surnames from the target field are recombined into author lists that do not correspond to any actual paper.
\item \textbf{Identifier fabrication and omission.} Under Non-Disclosure, DOI fields are frequently omitted or replaced with ``n/a''; in other conditions, models occasionally generate syntactically plausible but invalid DOIs.
\item \textbf{Title drift.} Generated titles are near-paraphrases of real papers---close enough to seem familiar but too different to match any indexed record above the 0.85 threshold.
\end{itemize}

These patterns explain why format compliance coexists with low verifiability: every field \emph{looks} correct in isolation, but the combination does not resolve to a real publication.

\subsection{Implications for Practice}

\paragraph{For tool and pipeline developers.}
Constraint satisfaction alone is not a reliable quality signal. Systems surfacing LLM-generated references should run post-hoc verification against multiple databases and treat \emph{Unresolved} as high risk.

\paragraph{For researchers using LLM writing assistance.}
Treat generated citations as candidates: require persistent identifiers (DOI/arXiv), cross-check metadata (Crossref/Semantic~Scholar, plus DBLP/OpenAlex for CS), and manually resolve remaining Unresolved before inclusion.

\paragraph{For the SE community.}
SE~\&~CS citations appear broadly comparable to other domains in our sample. Since Survey-style prompting---closest to SLR drafting~\cite{kitchenham2007slr}---is associated with a larger proprietary--open-weight gap, SE teams relying on open-weight models should be especially cautious.

\subsection{Threats to Validity}

\paragraph{Internal validity.}
Our label boundaries (0.85/0.60) and scoring weights are design choices applied uniformly; alternative settings could move citations between categories. We mitigate this with a manual audit (75\% agreement; $\kappa{=}0.63$) conducted by the authors. We tested at temperature zero with one prompt template per condition and one fixed phrasing per claim; results therefore characterize constraint \emph{types} rather than paraphrase robustness, and citation outcomes may vary under alternative wordings of the same claim.

\paragraph{External validity.}
{\tolerance=1200 Our 144 English-language claims may lack power for fine-grained comparisons (e.g., individual subdomains), and findings may not generalize to non-English domains or discipline-specific citation norms. With 24 claims per domain group, the domain-stratified analysis (Figure~\ref{fig:domain}) is descriptive; finer-grained breakdowns (e.g., SE subfields such as testing vs.\ requirements engineering) would require a substantially larger claim set. We evaluate four models at a single snapshot; the proprietary--open-weight gap could narrow as open-weight models scale up or incorporate retrieval, and newer releases may shift absolute rates. Results should therefore be read as characterizing structural patterns (constraint effects, proprietary--open-weight gaps) rather than as fixed benchmarks for any model version.\par}

\paragraph{Construct validity.}
Our pipeline relies on Crossref and Semantic Scholar; neither indexes everything, so some ``fabricated'' labels may be false positives (e.g., preprints, workshop papers, or regional-venue articles absent from both databases). Adding DBLP/OpenAlex would likely narrow the \emph{Unresolved} category. We also do not check whether a verified citation \emph{supports} the claim; citation--claim alignment is an important next step.

\subsection{Future Work}

Three directions extend this work beyond the limitations above. First, systematically varying claim phrasing (paraphrase robustness) would disentangle prompt sensitivity from constraint effects and clarify how much citation output depends on surface wording. Second, comparing closed-book generation against retrieval-augmented settings under the same constraint regimes would isolate how much hallucination stems from lack of grounding versus policy pressure. Third, embedding the verification pipeline as a real-time post-generation filter---for example, an IDE plugin or a CI check for manuscript drafts---would test its practical utility in SE writing workflows.

\section{Conclusion}

Across four models and five conditions, deployment constraints consistently worsen citation hallucination---but in qualitatively different ways. Temporal constraints sharply reduce verifiability while maintaining format compliance; Survey prompting widens the proprietary--open-weight gap; Non-Disclosure instructions shift errors from ``obviously wrong'' into ``hard to tell''; and models keep citing at high volume even as verifiability erodes. These patterns suggest that prompt engineering alone is unlikely to solve citation hallucination; reliable generation will require retrieval-augmented architectures, built-in verification, or both. For SE researchers, the practical implication is clear: any LLM-generated reference list should be treated as a draft requiring independent verification against scholarly databases before inclusion in reviews, technical reports, or tooling pipelines. More broadly, our benchmark and pipeline provide a reusable foundation for tracking whether citation reliability improves as new models and retrieval-augmented architectures emerge.

\paragraph{Data availability.}
A replication package (claim dataset, prompt templates, pipeline code, manual validation annotations, and raw results) is available at \url{https://github.com/Zerichen/Citation-Hallucination}.

\bibliographystyle{ACM-Reference-Format}
\bibliography{references}

\end{document}